\documentclass[amsmath,amssymb,aps,twocolumn,floats,superscriptaddress]{revtex4}

\usepackage{graphicx}
\usepackage{amssymb}
\usepackage{amsmath}
\usepackage{hyperref}
\usepackage{footnote}
\usepackage{color}
\usepackage{multirow}
\usepackage{enumerate}
\usepackage{soul}
\usepackage[FIGTOPCAP]{subfigure}
\pdfoutput=1 
\def \hH{ \hat{\mathcal{H}}}
\newcommand{\PD}{\phantom{\dag}}

\begin{document}
 
\title{Minimal models for nonreciprocal amplification using biharmonic drives}
\author{A. Kamal}
\affiliation{Research Laboratory of Electronics, Massachusetts Institute of Technology, Cambridge, MA 02139, USA}
\author{A. Metelmann}
\affiliation{Department of Electrical Engineering, Princeton University, Princeton, NJ 08544, USA}
 
\date{\today}
\begin{abstract}
We present a generic system of three harmonic modes coupled parametrically with a time-varying coupling modulated by a combination of two pump harmonics, and show how this system provides the minimal platform to realize nonreciprocal couplings that can lead to gainless photon circulation, and phase-preserving or phase-sensitive directional amplification. Explicit frequency-dependent calculations within this minimal paradigm highlight the separation of amplification and directionality bandwidths, generic to such schemes. We also study the influence of counter-rotating interactions that can adversely affect directionality and associated bandwidth; we find that these effects can be mitigated by suitably designing the properties of the auxiliary mode that plays the role of an engineered reservoir to the amplification mode space.
\end{abstract}
%
%
\maketitle
%
\section{Introduction}
%
Quantum-limited detectors and amplifiers are important modules for practical quantum information architectures. High-efficiency signal processing implemented with these systems has enabled single shot readout \cite{UCSBJeffrey2014} and real-time feedback control \cite{DelftRiste2013} of quantum bits in recent times. Realizing quantum-limited detection is intimately tied to the minimality of the mode space of amplification, as each additional mode introduced into the system potentially brings along its associated noise, the minimum being the quantum noise or zero-point fluctuations of the mode. Parametric systems achieve quantum-limited amplification by splitting a pump photon(s) between two channels, the (desired) signal and the (auxiliary) idler, a process which leads to the well-known quantum limit of half-a-photon of added noise at the signal frequency \cite{Caves1982}. Such amplification, in general, is described as a scattering between input and output signal and idler field amplitudes $a_{\rm sig,idl}$,
\begin{eqnarray*}
\left(\begin{array}{c}
a_{\rm sig}\\
a_{\rm idl}
\end{array}\right)^{\rm out}
=
\left(\begin{array}{cc}
\sqrt{\mathcal{G}+1} & \sqrt{\mathcal{G}} \\
\sqrt{\mathcal{G}} & \sqrt{\mathcal{G} + 1} 
\end{array}\right)
\left(\begin{array}{c}
a_{\rm sig}\\
a_{\rm idl}
\end{array}\right)^{\rm in}.
\end{eqnarray*}
where $\mathcal{G} > 1$ denotes the gain of the amplifier. As evident, such scattering is symmetric between signal and idler. Breaking this symmetry and realizing directional amplification is of immediate relevance to multiple quantum information processing (QIP) platforms, as it would (i) ensure unidirectional information transfer, (ii) prevent any noise impinging on the output port from getting amplified and re-directed to the signal-source (such as qubits), and (iii) significantly simplify measurement chains by eliminating the need of bulky components such as circulators and isolators, ultimately paving the way towards fully-integrated QIP. 
\par
It is worth noting that ideal directionality, while remaining strictly confined to two modes,
\begin{eqnarray*}
\left(\begin{array}{c}
a_{\rm sig}\\
a_{\rm idl}
\end{array}\right)^{\rm out}
=
\left(\begin{array}{cc}
0 & 0\\
\sqrt{\mathcal{G}} & \sqrt{\mathcal{G} + 1} 
\end{array}\right)
\left(\begin{array}{c}
a_{\rm sig}\\
a_{\rm idl}
\end{array}\right)^{\rm in},
\end{eqnarray*}
is forbidden by the requirement of symplectic structure of scattering \footnote{Symplectic symmetry follows from the bosonic commutation relations and the associated conserved quantity is the mode space of amplification.}. Thus the challenge is to realize nonreciprocal signal transfer and amplification while introducing a  minimum number of additional modes and preserve the quantum-limited operation of the device. Given the application potential of such systems, recent years have witnessed a strong surge in theoretical \cite{Koch2010, Kamal2011, Kamal2012, Ranzani2014, Metelmann2015, Kerckhoff2015} and experimental \cite{Abdo2013, Tzuang2014, Fleury2014, Sliwa2015} efforts that have aimed to realize quantum-limited nonreciprocity at acoustic, microwave and optical frequencies. 
\par
In this work, we analyze nonreciprocal photon dynamics in a framework that emphasizes minimality of the mode space and parametric pumping --- a feature especially desirable for hardware-efficient and scalable implementations of such detection protocols. Considering a generic system of parametrically-coupled three harmonic modes, which is the natural next step in increased mode complexity, we show how a two-pump biharmonic drive of the form
\begin{equation}
	G(t) = G_{\omega_{p}} \cos(\omega_{p}t) + G_{2\omega_{p}} \cos(2\omega_{p} t + \alpha),
\label{EqBipump}
\end{equation}
suffices to implement various kinds of nonreciprocal couplings. Such biharmonic drives (${\alpha \neq n\pi}$) are an economical way to realize time-asymmetric driving. This has been exploited to various ends previously, such as realizing noise-induced ratchet dynamics in Brownian motors \cite{Hanggi2005}, directed diffusion of cold atoms in optical lattices \cite{Schiavoni2003}, manipulation of fluxon transport in annular Josephson junctions \cite{Ustinov2004} and asymmetric driving of Landau-Zener-St\"{u}ckelberg-Majorana (LZSM) interferences in double quantum dots \cite{Forster2015} and superconducting qubits \cite{Gustavsson2013}. An additional advantage associated with using biharmonic drives is their autonomous generation in nonlinear optical crystals \cite{SHG} and Josephson junctions in the voltage state \cite{Kamal2014}. 
\par
The paper is organized as follows: in section \ref{sec_Raman} we describe directional phase-preserving amplification realizable in a new class of amplifiers, which we call biharmonic Raman amplifiers. We present calculations for both unresolved and resolved sideband regimes, and compare the available directionality with each under inclusion of relevant frequency-dependent non-resonant corrections. In section \ref{sec_phasesensitive} we describe directional phase-sensitive amplification with a biharmonically-pumped three-mode system. In section \ref{sec_discussion}, we discuss generic behavior and tradeoffs concerning gain, bandwidth and directionality. We also establish the connection of biharmonic pumping schemes to recently proposed dissipation engineering frameworks \cite{Metelmann2015}, and show that the general recipe of balancing dissipative and coherent interactions for implementing nonreciprocity simply maps to tuning the amplitude ratio ($G_{2 \omega_{p}}/G_{\omega_{p}}$) and phase difference ($\alpha$) of the two
harmonics. We conclude with a summary of our results in section \ref{sec_conclusions}. Additional details are included in appendices \ref{app_SRA} and \ref{app_circ}.
%
%
\section{Directional phase-preserving amplification: Biharmonic Raman Amplifiers}
\label{sec_Raman}
%
%
Phase-preserving amplification refers to equal amplification of both quadratures of a photonic field; this process maintains the phase information of the amplified signal in the quadrature space. Our general scheme to realize a directional phase-preserving amplifier is best understood in the framework of stimulated Raman scattering. It involves using a pump tone blue-detuned from the lower sideband resulting in Stokes scattering of the pump photons and red-detuned from the upper sideband leading to anti-Stokes scattering [cf. Fig.~\ref{Fig.:Raman}]. While the single-pump Stokes and anti-Stokes scattering is symmetric, it has been shown previously that the addition of the second pump harmonic induces asymmetry in these conversion processes \cite{Kamal2014}. Here we elaborate how such a process leads to directional amplification in the reduced subspace of the two sidebands. The basic idea relies on balancing (or `interfering') an indirect interaction between the two sidebands mediated by the first pump harmonic (through a third auxiliary mode), with a direct coherent interaction between the sidebands mediated by the second pump harmonic. The indirect interaction mediated by the auxiliary mode models a dissipative interaction as discussed later in Sec. \ref{sec_discussion}.
\par
We discuss two different regimes of these amplifiers: 
\begin{enumerate}[(i)]
  \setlength{\itemsep}{0pt}
\item Unresolved sideband amplification, i.e. $\omega_{a} \ll  \kappa_{b}$, where $\kappa_{b}$ denotes the linewidth of the high frequency oscillator at  $\omega_{b}$ 
      and $\omega_a$ corresponds to the resonant frequency of the auxiliary mode [cf. Fig.~\ref{Fig.:Raman}(a)]. This regime corresponds to \emph{degenerate phase-preserving amplification} since both input and output channels are accessed through a single mode \cite{DevoretRoy2016}.
\item Resolved sideband amplification, i.e. $\omega_{a} \gg \kappa_{b}$. This regime corresponds to \emph{non-degenerate phase-preserving amplification} since input and output channels are accessed via two distinct modes
      [cf. Fig.~\ref{Fig.:Raman}(b) ].
\end{enumerate}
%
%
\subsection{Unresolved sideband (USB) amplification}
%
%
\begin{figure} 
  \centering\includegraphics[width=\columnwidth]{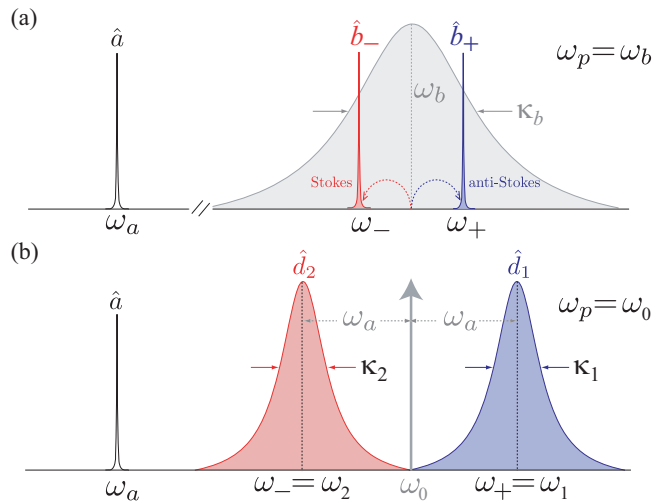} 
 	\caption{Frequency landscape of Raman amplifiers in (a) unresolved sideband regime ($\kappa_{a} \ll \omega_{a} < \kappa_{b}$), and (b) resolved sideband regime ($\kappa_{a} \ll \kappa_{1,2} < \omega_{a}$).  In both the cases, the presence of an additional tone at $2\omega_{p}$ leads to directional amplification between $\omega_{-} (\omega_{2})$ and $\omega_{+} (\omega_{1})$. }
 	\label{Fig.:Raman}
\end{figure}
%
We start with a generic Hamiltonian describing three harmonic modes coupled via time-varying (pairwise) interactions of the form,
\begin{align} \label{Eq.:3modeHamiltonian}
 \hH_{s} =&  
	   G_{1} (t) \left[\hat a + \hat a^{\dag} \right] \left(\hat d_1 + \hat d_1^{\dag} \right) 
           +  G_2(t) \left(\hat d_1 + \hat d_1^{\dag} \right)  \left(\hat d_2 + \hat d_2^{\dag} \right)
\nonumber \\ &
  	+  G_3(t) \left[\hat a + \hat a^{\dag} \right] \left(\hat d_2 + \hat d_2^{\dag} \right),
\end{align}
where $\hat a (\hat d_{1,2}) $ denotes the photon annihilation operator of the mode with frequency $\omega_a (\omega_{1,2})$. The modulations $G_{j}(t); (j \in 1; 2; 3)$ include the first and/or the second harmonic of an external pump at frequency $\omega_{p}$, cf. Eq.~(\ref{EqBipump}).
\par
We first consider the scheme depicted in Fig.~\ref{Fig.:Raman}(a). This system, in principle, is a two-mode system and can be realized using two parametrically coupled oscillators with frequencies $\omega_{a,b}$. In the unresolved sideband limit, where $\omega_{a}/\kappa_{b} \ll1$, we can consider the low-frequency mode at $\omega_{a}$ and the two sidebands at $\omega_{\pm} = \omega_{b} \pm \omega_{a}$ forming an effective three-mode system. Treating the sidebands as independent modes $\hat{b}_{\pm}$, we can map the system to the general three mode interaction Hamiltonian of Eq.~\ref{Eq.:3modeHamiltonian} via the correspondence
\begin{align}\label{Eq.MappingUSB}
 \hat d_1 \ \rightarrow \ \hat b_{+} = \hat b e^{ i\omega_a t },
\hspace{0.5cm}
 \hat d_2 \ \rightarrow \ \hat b_{-} = \hat b e^{-i\omega_a t },
\end{align} 
In the presence of a biharmonic drive of the form indicated in Eq.~(\ref{EqBipump}) and with $\omega_p = \omega_b$, the first pump harmonic induces Stokes (anti-Stokes) scattering to the lower (upper) sideband, while the second pump mediates an amplifying interaction between the two sidebands since $2\omega_{p} =\omega_{+} + \omega_{-}$. This leads to an effective mixing Hamiltonian in the interaction picture, with respect to the free Hamitonian, as
\begin{align}\label{EqHamil}
	\hH_{s} = &  \; G_{1}  \left( \hat{a} e^{-i \omega_{a} t} + \hat{a}^{\dagger} e^{i \omega_{a}t} \right)
			     \left(\hat{b} + \hat{b}^{\dagger} \right)
 \nonumber\\ &  \qquad   +  
	              \frac{G_{2}}{2} \left( \hat{b}\hat{b} \ e^{i\alpha} + \hat{b}^{\dagger}\hat{b}^{\dagger} \ e^{-i\alpha} \right).
\end{align}
The first line in Eq.~(\ref{EqHamil}) describes up- and downconversion processes between the low frequency mode $\hat a$ and the sidebands of the high frequency mode at $\omega_{\pm} = \omega_b \pm \omega_a$. The second line corresponds to an additional mixing pathway between these sidebands.  A combination of both these interactions results in asymmetric frequency conversion between auxiliary mode and sidebands \cite{Kamal2014}. Moreover, it allows for directional amplification within the sidebands. It is worthwhile to note here, that for $G_2 = 0$ and $\omega_P = \omega_b$, i.e., a monochromatic driving on resonance with the high frequency mode, the system realizes an effective two-mode phase-preserving amplifier which is not directional, but still has the interesting property of having no gain-bandwidth limitation.
The monochromatic driving scheme is closely related to a kind of dissipative amplifier introduced recently \cite{Metelmann2014} (see appendix \ref{app_SRA} for further details).
\par
Setting the coupling strengths of the two oscillators to the external input/output ports as $\kappa_{a,b}$, we can use standard input-output theory \cite{IOT} to derive the Heisenberg-Langevin equations describing the dynamics of our system. To highlight the relevant features of this setup, we start by focusing on the zero-frequency case. Thus the low frequency mode is on resonance and has the stationary solution,
\begin{align}
    \hat a[0] =& 
		- i \frac{2 G_1}{\kappa_a} \left( \hat b[ \omega_a]  +\hat b^{\dag}[ \omega_a] \right),
\label{Eqao}
\end{align}
where, we neglect any noise contribution driving the low frequency oscillator for simplicity. We see that the mode $\hat a$ is coupled to the two sideband lying at $\pm \omega_a$ in this rotated frame. Correspondingly, the sidebands couple to $\hat a[0]$ and $\hat a^{\dag}[2 \omega_a]$. For $\omega_a \gg \kappa_a$, the processes mixing Stokes and anti-Stokes with $\hat a^{\dag}[2 \omega_a]$ can be ignored under a rotating wave approximation in the $\hat{b}_{\pm}$ basis \footnote{This is equivalent to treating the two sidebands as independent modes.}. In this limit, we can use the stationary solution for $\hat a[0]$ to obtain the corresponding solutions for the sidebands
\begin{align}  \label{Eq.:EoMstationarySidebandNDPA}
 \chi_{+}^{-1}  \hat b[ \omega_a] =& -    \frac{2}{\sqrt{\kappa_b}}  \hat b_{\rm in}[  \omega_a]   
			- \left[ \mathcal C  + i \frac{2 G_2 }{\kappa_b} e^{-i \alpha} \right] \hat b^{\dag}[ \omega_a],
\nonumber \\  
 \chi_{-}^{-1} \hat b^{\dag}[ \omega_a] =& - \frac{2}{\sqrt{\kappa_b}} \hat b_{\rm in}^{\dag}[\omega_a]   
			+   \left[ \mathcal C   +   i \frac{2 G_2 }{\kappa_b} e^{i \alpha} \right] \hat b[\omega_a],
\end{align}
with the modified susceptibilities  $ \chi_{\pm}^{-1} =  \left[1 \pm \mathcal C  -i \frac{2 \omega_a}{\kappa_b}   \right] $ and the cooperativity $\mathcal C = \frac{4 G_1^2}{\kappa_a \kappa_b}$. The operators $\hat b_{\rm in}^{(\dag)} $  describe any input impinging at the sideband frequencies of the b-mode, i.e., an input signal one wishes to amplify or just thermal and vacuum fluctuations. Unlike the case of a single pump ($G_{2} =0$), there are differences in how each sideband couples to the other as reflected by the asymmetries in the terms within square brackets of Eq.~(\ref{Eq.:EoMstationarySidebandNDPA}). Moreover, this asymmetry is tunable with phase $ \alpha$ and the strength $G_{2}$ of the second pump. It is straightforward to see that for
\begin{align}\label{Eq.:DirCondNDPA}
  G_2  =  \frac{\kappa_b}{2} \mathcal C\;,
\hspace{0.5cm}
         \alpha =   - \frac{\pi}{2},
\end{align}
$ \hat b[\omega_a]$ decouples from the reduced system of the two sidebands. This realizes a directional interaction, as we now have the situation that $\hat b^{\dag}[\omega_a]$ is influenced by $\hat b  [\omega_a]$ but not vice versa. 
\par
In order to calculate the full nonreciprocal scattering matrix of the system, we use the standard input-output relation ${\hat o_{\rm out} = \hat o_{\rm in} + \sqrt{\kappa_o} \hat o, \ (o \in a,b)}$. Hereby we include the fluctuations impinging on the low frequency mode ($\hat a_{\rm in}$) as well, which we had neglected earlier. Then the zero-frequency scattering matrix in the basis $\mathbf{\hat D}[0] = [\hat a^{\PD}[0], \hat b^{\PD}[\omega_a],\hat b^{\dag}[\omega_a]]^{\rm T}$ becomes (for $\omega_{a}/\kappa_{b} \ll1$)
\begin{equation}  \label{Eq.:SmatrixSidebandNDPA}
 \mathbf{s}[0] =
\left(
\begin{array}{ccc}
	  \displaystyle     -1
	& \displaystyle     
			    \frac{ i 2 \sqrt{\mathcal C}}
				 {1   - \mathcal C}
	& \displaystyle     
			    \frac{ i 2 \sqrt{\mathcal C}}
			         {1   - \mathcal C}
\\[0.2cm]
	  \displaystyle     \frac{ i 2 \sqrt{\mathcal C}}
				 {1 + \mathcal C} 
	& \displaystyle       -\frac{1 - \mathcal C}
				 {1 + \mathcal C}
	& \displaystyle     0
\\[0.2cm]
	  \displaystyle     \frac{ -i 2 \sqrt{\mathcal C}}
				 {1 +  \mathcal C }
	& \displaystyle    - \frac{ 4 \mathcal C}
				 {1-  \mathcal C^2}
	& \displaystyle    - \frac{1 +  \mathcal C}
				 {1 -  \mathcal C}
\\
\end{array}
\right),
\end{equation} 
where $\mathbf{\hat D}_{\rm out}[0] = \mathbf{s}[0]  \mathbf{\hat D}_{\rm in}[0]$. The diagonal elements of this matrix correspond to the reflection coefficients, the off-diagonal elements $s_{21},s_{31}$ describe upconversion from the low frequency to the sideband frequencies while $s_{12},s_{13}$ describe the corresponding down-conversion. The important elements are $s_{23}$ and $s_{32}$ which describe the amplification between the two sidebands. An input signal at the upper sideband shows up amplified at the lower sideband, i.e., it gets down-converted in frequency, while any input on the lower sideband will never show up at the higher sideband as $s_{23} = 0$. Note, that for $\alpha =   \frac{\pi}{2}$ the  situation is reversed, i.e., a signal is  up-converted between the sidebands and amplified, but this leads to unwanted amplification of the reflected input signal too. 
\par
The zero-frequency gain can be read off from Eq.~(\ref{Eq.:SmatrixSidebandNDPA}) as
\begin{align}\label{Eq.:ZeroFrequGain}
 \mathcal G_{0}  \equiv \left| s_{32} [0]\right|^2 =&  \frac{ 16 \mathcal C^2  }{   \left[ \mathcal C^2 - 1 \right]^2},
\end{align}
which increases as $\mathcal C \rightarrow 1^{-}$; as usual stability requires $\mathcal C < 1 $.  From the scattering matrix in Eq.~(\ref{Eq.:SmatrixSidebandNDPA}) we see that in the large gain limit the output at the upper sideband contains noise stemming from the low frequency auxiliary mode alone; also the reflection at the input vanishes ($s_{22} =0$), a feature desirable for applications such as qubit readout. Moreover, this amplification process is quantum-limited, as can be seen by calculating the added noise, 
\begin{align}
\bar n_{\rm add}   =&  \frac{1}{2} +\bar n_{b}^{T}
		     +   \left(2\bar{n}_{b}^{T} + \bar n_{a}^{T} + \frac{3}{2} \right) \frac{1}{\mathcal G_0} + \mathcal{O}\left(\frac{1}{\mathcal{G}_{0}^{2}}\right).
\end{align}
In the regime of large gain ($\mathcal G_0$) and zero-temperature baths ($ n_{a}^{T} = n_{b}^{T} \approx 0$), we obtain the quantum limit of $\bar n_{\rm add}= 1/2$.
%
%
\subsubsection*{Frequency dependence of directional gain}
\begin{figure}[t!] 
  \centering\includegraphics[width=0.95\columnwidth]{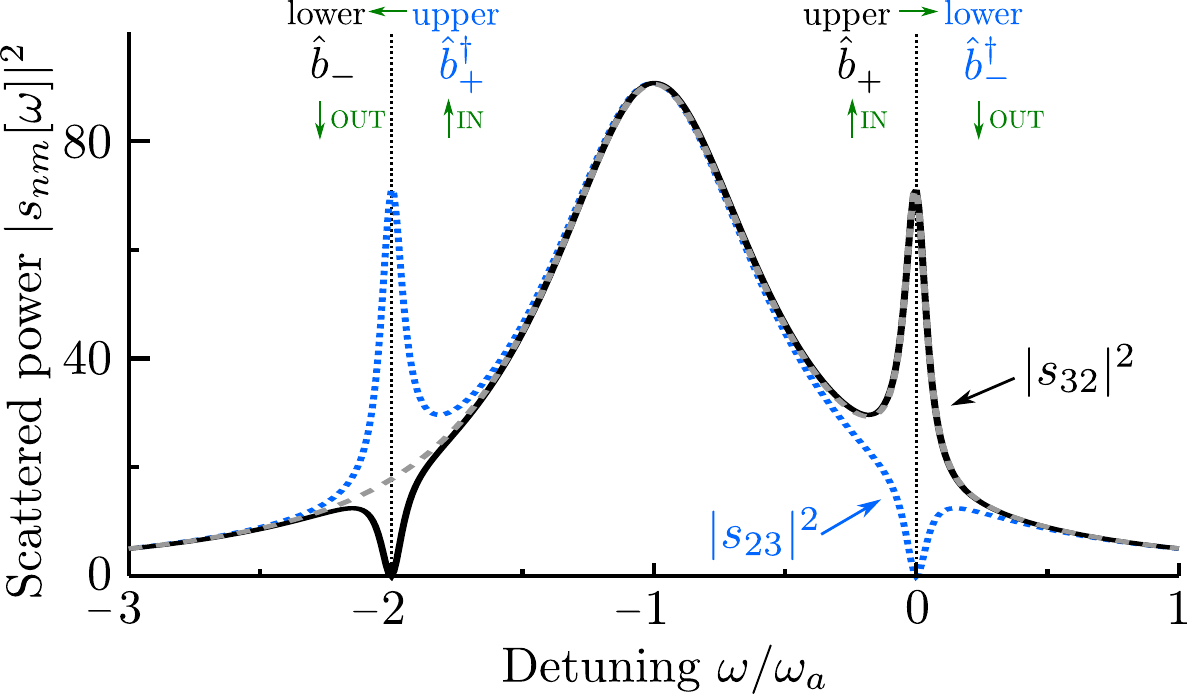} 
 	\caption{Frequency dependence for the scattering matrix elements of the unresolved sideband amplification. 
 	         The dashed-gray line depicts the gain $|s_{32}[\omega]|$ under RWA, cf. Eq.~(\ref{Eq.:GainSidebandNDPA}).
 	         The non-RWA results for forward and reverse gains are plotted as the solid-black and the dotted-blue lines
 	         respectively. The scattering matrix elements at the sideband resonances $\omega_{b}\pm\omega_a$ describe 
 	         downconversion to the lower sideband for an input signal injected at the upper sideband.  
 	         Parameters used in the calculation are $\mathcal  C = 0.9$, $\omega_a/\kappa_b = 0.1$ , $Q_a = 10$.}
 	\label{Fig:SmatNDPA}
\end{figure}

%
Finally, we take a look at the expressions for the gain and the reverse gain as a function of frequency. We still consider the situation where we have an input signal at the upper sideband which gets amplified and completely down-converted to the lower sideband under the directionality condition Eq.~(\ref{Eq.:DirCondNDPA}). The finite frequency gain is ($\omega_a \ll \kappa_a$)
\begin{align}\label{Eq.:GainSidebandNDPA}
\mathcal G[\omega] =&  \frac{ 16 \mathcal C^2  \left[1 + \frac{\omega^2}{\kappa_a^2}    \right] \left[ 1 + \frac{4 \omega^2}{\kappa_a^2} \right]^{-1}}
				{\left[ \left[\mathcal C -  1 \right]^2 +   \frac{4\left(\omega + \omega _a   \right)^2}{ \kappa _b^2}    \right]
				\left[ \left[\mathcal C +  1 \right]^2 +   \frac{4\left(\omega + \omega _a   \right)^2}{ \kappa _b^2}    \right]},
\end{align}
where we keep the ratio of $\omega_a/\kappa_b$ unspecified, i.e., we do not restrict ourselves to unresolved sideband regime. From the full expression for the frequency-dependent gain, we see that the gain profile shows a peak at ${\omega = - \omega_a}$ which corresponds to the resonance frequency $\omega_b$ in this rotated frame [cf. Fig.~\ref{Fig:SmatNDPA}].
%
%
%
%
However, the reverse gain only vanishes at the lower sideband $(\omega = 0)$,
\begin{align}\label{Eq.:RevGainSidebandNDPA}
\bar{\mathcal G}[\omega] \equiv \left|s_{23}[\omega] \right|^2 
			= \frac{ \frac{ \omega^2}{\kappa_a^2}}{ \left(1 + \frac{\omega^2}{\kappa_a^2}    \right)} \mathcal G[\omega],
\end{align}
which describes the up-conversion of possible inputs, i.e., thermal or vacuum fluctuations from the lower sideband. The ideal situation corresponds to a vanishing of this reverse gain over a wide frequency bandwidth. The directionality bandwidth scales with $\kappa_a$, as evident from Eq.~(\ref{Eq.:RevGainSidebandNDPA}). However, in order to treat both sidebands independently we need to have $\kappa_{a} \ll \omega_{a}$, i.e. a high-quality factor for low frequency $\hat{a}$ mode. One would think that having this mode at high frequency may do the trick and a large directionality bandwidth could, in principle, be maintained. However, just having a large $\omega_a$ is not sufficient, as for $\omega_a \rightarrow \infty $ the gain vanishes as well. The relevant quantity here is the ratio $\omega_a/\kappa_b$; this becomes obvious if we consider the gain at resonance [cf. Eq.~(\ref{Eq.:GainSidebandNDPA})] for $\mathcal C \rightarrow 1^{-}$,
\begin{align}
\mathcal G[0]  =&   \frac{1}{\frac{ \omega_a^2}{ \kappa_b^2}     \left( 1 +  \frac{ \omega_a^2}{ \kappa_b^2}    \right)} \approx \frac{ \kappa_b^2}{\omega_a^2}.
\label{Eq.ZerogainUSB}
\end{align} 
The gain saturates and the maximal gain value scales with $(\kappa_b/\omega_a)^2$ for $\mathcal C \rightarrow 1^{-}$. Thus, the unresolved sideband regime is an important ingredient to obtain any gain at all in this scheme. 
%
%
\subsection{Resolved sideband (RSB) amplification}
\label{Sec.:3modeNDPA}
%
Our analysis in the previous section showed that the parameter hierarchy $\kappa_a \ll \omega_a < \kappa_b$ is crucial to realize a directional phase-preserving amplification between the sidebands. The restriction to the unresolved sideband regime, however, constrains both the forward gain and directionality of such a biharmonic Raman amplifier. For instance a $20$~dB of gain would already require a ratio $\kappa_b/\omega_a \simeq 10$; though achievable in opto/electro-mechanical setups, this limits the application potential of such a scheme in superconducting setups employing microwave frequencies. In this section, we show how operating in the resolved sideband regime alleviates these difficulties. We now extend our system to include three independent oscillators with resonance frequencies $\omega_a \ll \omega_2 < \omega_1$. As shown in Fig.~\ref{Fig.:Raman}(b), the independent oscillator modes at frequencies $\omega_{1,2}$ play the role of the sidebands of the USB amplifier case. Choosing the driving frequencies $\omega_{P,i},(n \in (1,2,3)) $ of the time-dependent couplings $G_{j}(t)$ in Eq.~(\ref{Eq.:3modeHamiltonian}) as
\begin{subequations}
\begin{align}
 \omega_{P,1} =& \ \omega_1 - \omega_a  \equiv \omega_0,
 \\
\omega_{P,2} = & \ \omega_2 + \omega_a  \equiv \omega_0,
\\
\omega_{P,3} = & \ \omega_1 + \omega_2   \equiv 2   \omega_0,
\end{align}
\end{subequations}
makes the following interactions resonant in the system, just as in the unresolved sideband regime:
\begin{enumerate}[(i)]
  \setlength{\itemsep}{0pt}
  \setlength{\parskip}{0pt} 
\item the first harmonic at $\omega_0$ mediates a hopping interaction between the auxiliary mode $\hat a$ and mode $\hat d_1$
\item the first harmonic at $\omega_0$ mediates an amplifier interaction between the auxiliary mode $\hat a$ and mode $\hat d_2$
\item the second harmonic at $2\omega_0$ mediates an amplifier interaction between the modes $\hat d_2$ and $\hat d_1$.
\end{enumerate}
As before, we work in an interaction picture with respect to the oscillators' free Hamiltonian and obtain   
\begin{align} 
 \hH =&    G_1 \left( \hat a \hat d_1^{\dag}  + \hat a^{\dag}  \hat d_2^{\dag} \right)  
	+  G_2  \hat d_1^{\dag} \hat d_2^{\dag} e^{- i \alpha} + \hH_{\rm CR} + h.c.,
\label{Eq.:Ham3mode}
\end{align}
where $\hH_{\rm CR}$ contains the counter-rotating terms.
\par
We consider the same situation as before, i.e., the directional amplification of an input signal at the upper mode with frequency $\omega_1$ which shows up at the output of the mode at frequency $(\omega_2)$. Using input-output theory and utilizing the directionality condition of Eq.~(\ref{Eq.:DirCondNDPA}), we obtain the same zero-frequency scattering matrix as in Eq.~(\ref{Eq.:SmatrixSidebandNDPA}) but now in the basis  $\mathbf{\hat D}[0] = [\hat a^{\PD}[0], \hat d_{1}^{\PD}[0],\hat d_{2}^{\dag}[0]]^{\rm T}$. The finite frequency gain for RSB amplifier is given as (neglecting $\hH_{\rm CR}$)
\begin{align}\label{Eq.:Gain3modeNDPA}
\mathcal G[\omega]   =&  
			\frac{16 \mathcal C ^2   \left(1+ \frac{\omega^2}{\kappa_a^2}\right)}
			{\left[1+ \frac{4 \omega^2}{\kappa_a^2}\right]
			 \left[ \left(\mathcal C - 1\right)^2 + \frac{4\omega^2}{\kappa_b^2} \right] 
			 \left[ \left(\mathcal C + 1\right)^2 + \frac{4\omega^2}{\kappa_b^2} \right]  },
\end{align}
where, for simplicity, we have assumed equal decay rates for both the oscillator modes 1 and 2, $\kappa_{1} =\kappa_{2} \equiv \kappa_{b}$.  The reverse gain of RSB amplifier is the same as that given in Eq.~(\ref{Eq.:RevGainSidebandNDPA}) with $\mathcal{G} [\omega]$ now given by Eq.~(\ref{Eq.:Gain3modeNDPA}). Thus we have the same situation as in the USB case; the reverse gain vanishes on resonance and the directionality bandwidth increases with $\kappa_a$. The important difference, however, is that the gain in Eq.~(\ref{Eq.:Gain3modeNDPA}) does not depend on the ratio $\omega_a/\kappa_b$, in contrast to Eq.~(\ref{Eq.:GainSidebandNDPA}). Thus, there is no saturation of the gain as encountered in the unresolved sideband regime [cf. Eq.~(\ref{Eq.ZerogainUSB})]. 
\par
We discuss another mode of operation of this system, namely a nonreciprocal photon transmission without gain or a frequency circulator, in appendix \ref{app_circ}.
%
%
\subsection{Influence of counter-rotating terms: USB versus RSB}
%
All the calculations in previous sections were done neglecting the counter-rotating terms. These terms oscillate with twice of one of the systems frequencies, i.e., $e^{i 2\omega_{j} t}, (j \in a,1,2,0)$; since the smallest frequency is $\omega_a$, the terms oscillating at $2\omega_{a}$ are the most relevant counter-rotating terms. Under this assumption, we describe the counter-rotation Hamiltonian as
\begin{align} 
\hH_{\rm CR} \simeq&   
		G_1 \left(  \hat a^{\dag}  \hat d_1^{\dag} e^{ i 2 \omega_a   t }   +   \hat a \hat d_2^{\dag}   e^{-i 2  \omega_a   t }   \right)  + {\rm h.c.},
\label{Eq.:Hamilcounter}
\end{align}
%
%
%
\begin{figure}[t!] 
\centering
 \includegraphics[width=\columnwidth]{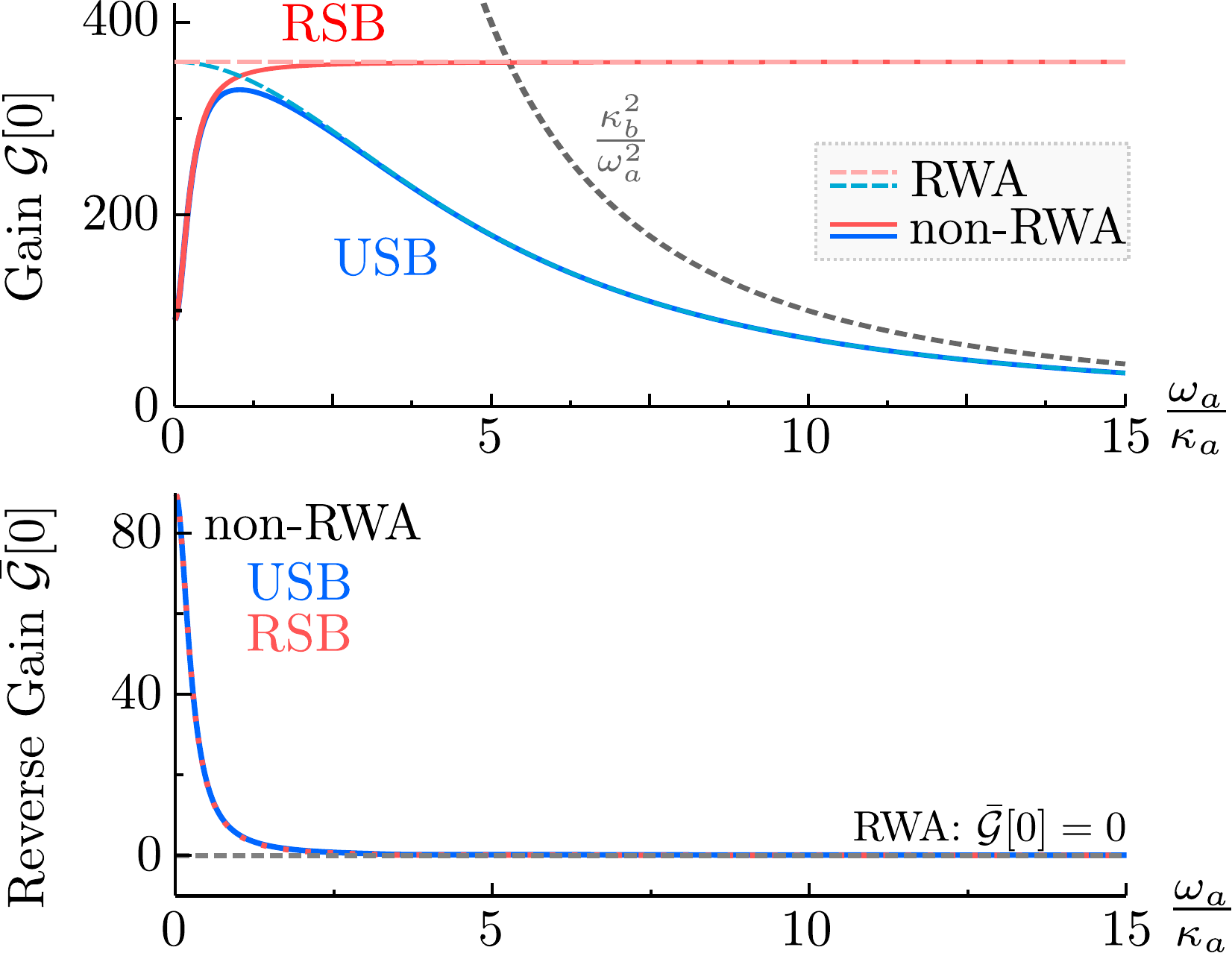}
	\caption{Comparison of USB and RSB biharmonic Raman amplifiers. Parameters are $\mathcal C = 0.9$ and $\kappa_b = 100 \kappa_a$. The upper (lower) graph depicts the forward (reverse) gain at resonance, as a function of quality factor of the auxiliary mode $Q_{a} =\omega_{a}/\kappa_{a}$. In the USB case, the gain first improves as $Q_{a}$ increases and then decreases and reaches the limit $\kappa_b^2/\omega_a^2$ (grey dashed line) for large $Q_{a}$. The RSB amplifier, on the other hand, suffers from no gain saturation and it operates at the expected maximal gain value, i.e., $\mathcal G \sim   \mathcal G_0 \approx 360$ for high $Q_{a}$. Further, a comparison of the RWA and non-RWA results shows that $\omega_a > \kappa_a$ is necessary to suppress reverse gain in both the USB and RSB schemes (under RWA, the reverse gain is always zero).}
\label{Fig.:ComparisonRaman}
\end{figure}
%
\par
Fig.~\ref{Fig.:ComparisonRaman} compares the forward and reverse gains as a function of $Q_{a}$, calculated including the effect of Eq.~(\ref{Eq.:Hamilcounter}) for both the USB- and RSB types of biharmonic Raman amplifiers. It is clear that the counter-rotating terms lower the forward gain of both amplifiers unless filtered out by a sufficiently high-$Q_{a}$ for the auxiliary mode,$\omega_a/\kappa_a \gg 1$. Furthermore, though (reduced) forward gain persists in low-$Q_{a}$ regime, the reverse gain vanishes and directionality is restored only in the high-$Q_{a}$ limit. 
\par
While the behaviors of the RSB amplifier and the USB amplifier coincide in the low-$Q_{a}$ regime, the gain for the USB amplifier decreases strongly with increasing $Q_{a}$. The saturation of the gain sets in as $\omega_{a}/\kappa_{a} \rightarrow \kappa_{b}/\kappa_{a}$ [cf. Eq.~(13)]. This significantly limits the useful bandwidth over which it can be exploited as a \emph{directional} phase-preserving amplifier. Operating in the RSB regime alleviates this problem and drastically increases the bandwidth over which the amplifier is directional, though it still requires a modest $Q_{a}$ for the auxiliary mode. 
%
%
\section{Directional phase-sensitive amplification}
\label{sec_phasesensitive}
\begin{figure}[ht]
\centering
 \includegraphics[width=\columnwidth]{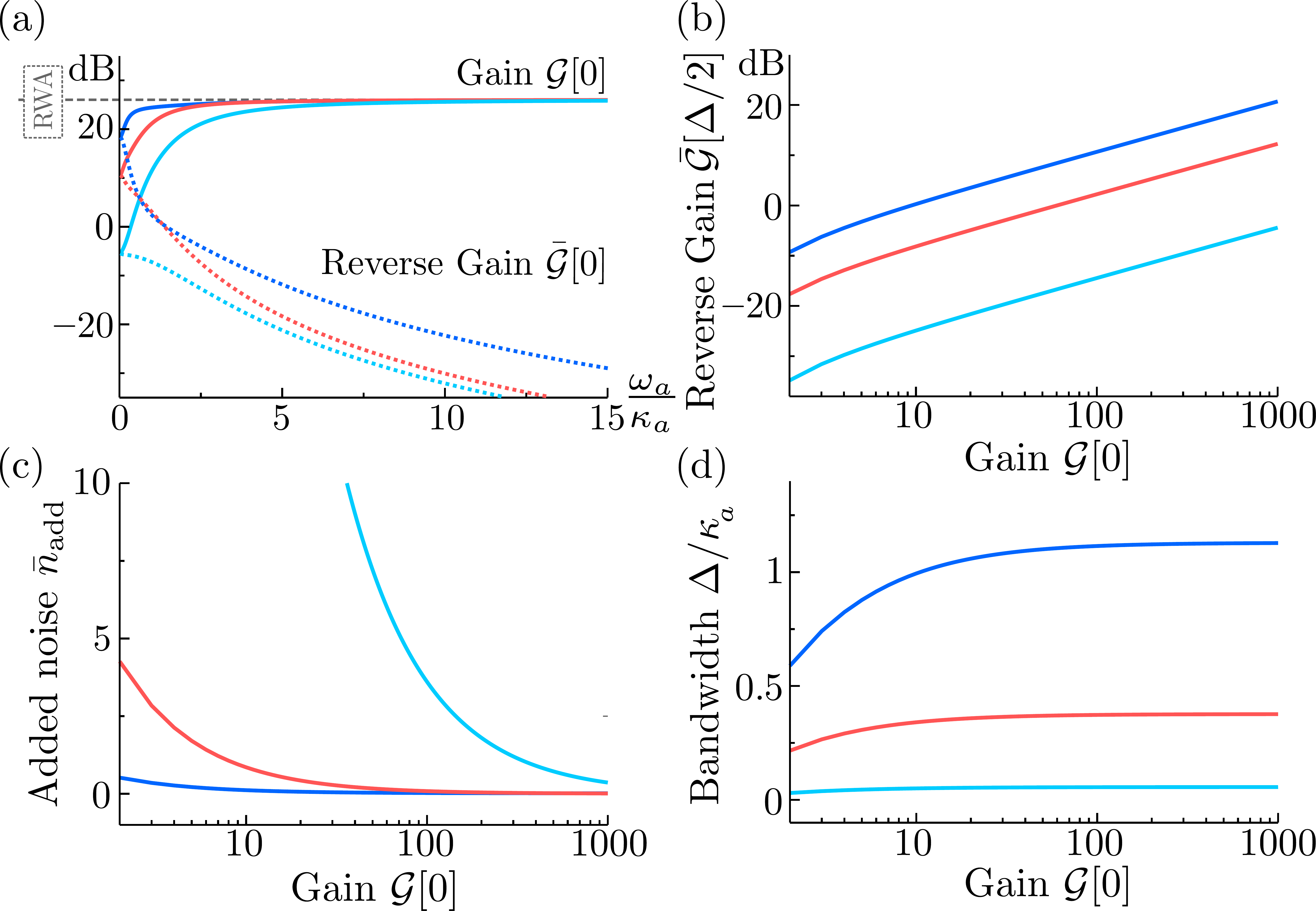}
	\caption{Characteristics for the directional  phase-sensitive amplifier scheme, for three different ratios of $\kappa_{a}/\kappa$ = 0.1 (blue), 0.5 (red), 0.9 (cyan). For each plot, the calculation assumed directionality condition namely $G_3 = \kappa_a/2$. (a) Frequency-dependent forward and reverse gains as a function of $Q_a = \omega_a/\kappa_a$ for the auxiliary mode. The dashed gray line shows the gain without the counter rotating terms which is independent from $Q_a$, here $\mathcal G[0]^{\textup{RWA}}$ = 26 dB at resonance,  while the corresponding reverse gain is always zero. We see that for moderate quality factors the reverse gain is highly suppressed, while the gain approaches the RWA result; hence we chose $\omega_a/\kappa_a = 10$ for the remaining graphs (b-d).
        (b) Reverse gain evaluated at half of the amplification bandwidth, i.e., at $\omega = \Delta/2$ with $\Delta$ being the full-width at half-maximum of the forward gain profile. (c) Added noise quanta versus zero frequency gain for zero temperature baths. (d) Amplification bandwidth expressed in units of linewidth of auxiliary mode, $\Delta/\kappa_{a}$. Crucially, the bandwidth does not decrease while increasing the gain, hence we have no gain-bandwidth limitation.
}
\label{Fig.:DPSAchar}
\end{figure}

%
We now present a biharmonically-pumped three-mode scheme that realizes directional phase-sensitive amplification, i.e. only one of the input quadratures gets amplified and appears at the output. Originally proposed in Ref. \cite{Metelmann2015}, the key idea here is to realize a quantum non-demolition (QND) interaction between the input and output modes. In the most general case, at least six driving tones are required to mediate the requisite interactions. In this section, we show how such an interaction can be realized using a biharmonic tone, with only a single constraint on the auxiliary mode frequency $\omega_{a}$ being degenerate with one of the input or output modes. To this end, we consider the most general three-mode Hamiltonian of Eq.~(\ref{Eq.:3modeHamiltonian}) under the pumping conditions,
\begin{subequations}
\begin{align}
& \omega_{1} + \omega_{a} \equiv 2           \omega_{0}, \quad \omega_{1} - \omega_{a} = \omega_{0}\\
& \omega_{1} + \omega_{2} \equiv 2           \omega_{0}, \quad \omega_{1} - \omega_{2} = \omega_{0}\\
& \omega_{2} + \omega_{a} \equiv \omega_{0}, \quad \omega_{2} - \omega_{a} = 0,
\end{align}
\end{subequations}
which selects the following interactions to be resonant:
\begin{enumerate}[(i)]
  \setlength{\itemsep}{0pt}
  \setlength{\parskip}{0pt} 
\item the first harmonic at $\omega_0$ mediates a hopping interaction while the second harmonic mediates an amplifying interaction between the auxiliary mode $\hat a$ and mode $\hat d_1$
\item the first harmonic at $\omega_0$ mediates a hopping interaction while the second harmonic mediates an amplifying interaction between modes $\hat{d}_{1}$ and $\hat d_2$
\item the first harmonic at $\omega_0$ mediates an amplifying interaction between the the auxiliary mode $\hat a$ and mode $\hat d_2$
\end{enumerate}
This pumping scheme can be easily realized for any combination of mode frequencies of the form $\omega_{a} = \omega_{2} = (1/3) \omega_{1}$. Note that the interaction between the auxiliary mode and mode 2 can only be of the amplifying type in this case since these modes are designed to be at the same frequency. 
%
%
\begin{figure}[t!]
\centering
 \includegraphics[width=0.8\columnwidth]{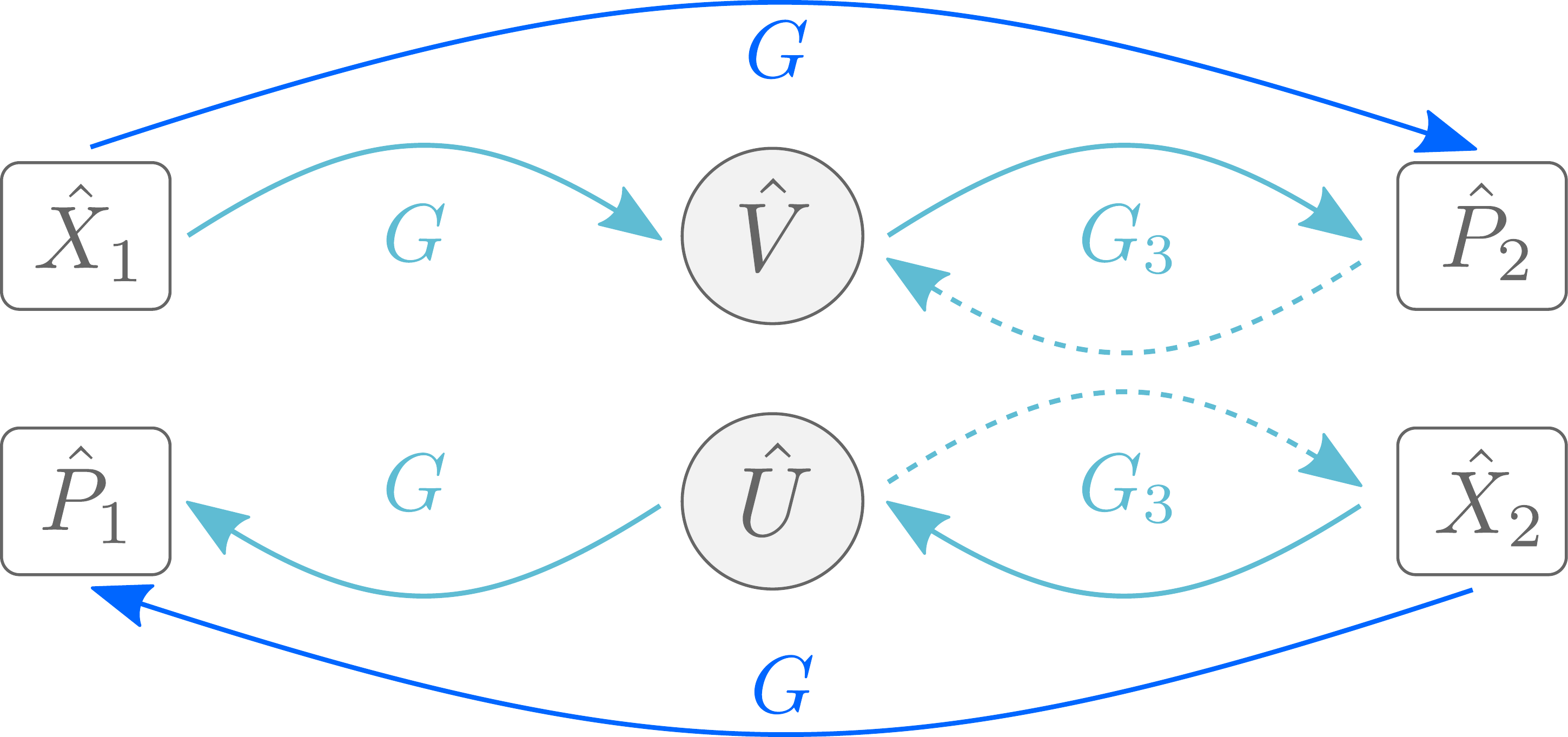}
	\caption{The direct interaction between the modes 1 and 2 (solid-blue) alone corresponds to an information transfer where the information of the X-quadrature of each mode is dumped into the  P-quadrature of the respective opposite mode. The coupling to the auxiliary mode breaks this reciprocal process by mediating the same transfer process between the modes (solid-cyan). Balancing these process leads to perfect decoupling of mode 1 from mode 2. The dashed lines show the feedback loops which are unavoidable in this minimal biharmonic driving scheme, however, they are not damaging for the whole scheme to work.}
\label{Fig.:DPSA}
\end{figure}
%
%
This leads to following interaction Hamiltonian (under RWA),
\begin{eqnarray}
\hH = G \hat{X}_{1}\hat{X}_{2} - G_{3} [\hat{X}_{2} \hat{V} + \hat{P}_{2}\hat{U}] + G \hat{X}_{1} \hat{U},
\label{EqHPSAquad}
\end{eqnarray}
with $\hat{X}_{i} = (\hat{d}_{i} + \hat{d}_{i}^{\dagger})/\sqrt{2}, \hat{P}_{i} = -i(\hat{d}_{i} - \hat{d}_{i}^{\dagger})/\sqrt{2}, \; (i \in 1,2)$ being the quadratures associated with input-output modes, and $(\hat{U}, \hat{V})$ the quadratures associated with the auxiliary mode. Here we have chosen $G_{1} = G_{2} \equiv G/2$ and phase difference between $G$ and $G_{3}$ to be $\pi/2$ [cf. Eq.~(\ref{Eq.:3modeHamiltonian})].  Eq.~(\ref{EqHPSAquad}) shows that $\hat{X}_{1}$ is a QND observable and is, therefore, preserved from quadrature mixing as desired for phase-sensitive amplification. In the optimal case $\hat{X}_{2}$ would be as well a QND-observable \cite{Metelmann2015}; this can be accomplished by balancing out the term  $\hat{P}_{2}\hat{U}$  either through a static coupling \cite{Abdo2013} between modes $\hat{a}$ and $\hat{d_{2}}$, or by lifting their degeneracy at the expense of introducing additional pumps. Figure~\ref{Fig.:DPSA} illustrates the information transfer mediated by the Hamiltonian in Eq.~(\ref{EqHPSAquad}). The information in the X-quadratures of each mode is transferred to the respective P-quadratures of the other mode in two ways: via a direct transfer and via a transfer over the auxiliary mode. Balancing these interactions as before allows us to realize desired unidirectional information transfer between selected quadratures. This can be easily seen from the zero-frequency scattering matrix calculated from the coupled equations of motion for the quadratures, which after the elimination of auxiliary mode reads
\begin{equation}  \label{Eq.:SmatrixSidebandDPA}
 \mathbf{s}[0] =
\left(
\begin{array}{cccc}
	  \displaystyle    -1
	& \displaystyle     
			    0 
	& \displaystyle     0
	& \displaystyle     0
\\[0.2cm]
	  \displaystyle     0 
	& \displaystyle    -1
	& \displaystyle     0
	& \displaystyle     0
\\[0.2cm]
	  \displaystyle     0
	& \displaystyle     0
	& \displaystyle    - \frac{\kappa +\kappa_{a}}{\kappa -\kappa_{a}}
	& \displaystyle     0
 \\[0.2cm]
	  \displaystyle     \frac{8G}{\kappa -\kappa_{a}}
	& \displaystyle     0
	& \displaystyle    0
	& \displaystyle     - \frac{\kappa +\kappa_{a}}{\kappa -\kappa_{a}}
\\
\end{array}
\right),
\end{equation} 
where $\mathbf{\hat D}_{\rm out} = \mathbf{s} \mathbf{\hat D}_{\rm in} + \mathbf{\xi}$, 
$\mathbf{\hat D}_{\rm i \in ( \rm {in, out} )} = \left[\hat{X}_{1}^{\rm i}, \hat{P}_{1}^{\rm i}, \hat{X}_{2}^{\rm i}, \hat{P}_{2}^{\rm i}\right]^T$, 
$\kappa_{1} = \kappa_{2} \equiv \kappa$, and $\mathbf{\xi}$ denotes the noise contribution from the auxiliary mode. Here we have set the interaction strength $G_{3}=\kappa_{a}/2$ to impose unidirectional coupling from mode 1 to mode 2.  This results in directional phase-sensitive amplification: the $\hat{P}_{2}^{\rm out}$ contains the amplified input quadrature $\hat{X}_{1}^{\rm in}$, while no input on either of the cavity-2 quadratures shows up at the output of cavity 1. The amplitude gain scales with $G$ and stability requires $\kappa_{a} \ll \kappa$ as the paramp interaction between the auxiliary mode and cavity-2 mode introduces anti-damping of the latter.  Moreover, the added noise 
\begin{eqnarray}
	\bar{n}_{\rm add} = \left(\frac{\kappa + \kappa_{a}}{8 G}\right)^{2}\left(\bar{n}_{2}^{T} + \frac{1}{2} \right) + \frac{\kappa_{a}\kappa}{16 G^{2}} \left(\bar{n}_{a}^{T} + \frac{1}{2} \right) 
\end{eqnarray}
goes to zero in the high-gain limit $G \gg \kappa,\kappa_{a}$ [cf. Fig.~\ref{Fig.:DPSAchar}(c)], as desired for ideal phase-sensitive amplification. The expressions for frequency-dependent forward gain and reverse gain,
\begin{eqnarray}	
	\mathcal{G} [\omega]  &=& |s_{P_{2} \leftarrow X_{1}}[\omega]|^{2} \nonumber\\
	& =& \frac{16 \mathcal{C}_{a} (1 + \frac{\omega^{2}}{\kappa_{a}^{2}})}{\left(1+ \frac{4 \omega^{2}}{\kappa_{a}^{2}}\right)\left(\left[\frac{4 \omega^{2}}{\kappa_{a}^{2}}+ \frac{\kappa}{\kappa_{a}} - 1\right]^{2} + \frac{4\omega^{2}}{\kappa_{a}^{2}}\left(1+\frac{\kappa}{\kappa_{a}}\right)^{2}\right)}, \nonumber\\
\label{Eq.:FwdGainPSA}\\
	\mathcal{\bar{G}} [\omega] &=& |s_{P_{1} \leftarrow X_{2}}[\omega]|^{2} = \frac{\frac{\omega^{2}}{\kappa_{a}^{2}}}{1 + \frac{\omega^{2}}{\kappa_{a}^{2}}}\mathcal{G} [\omega], 
\label{Eq.:RevGainPSA}
\end{eqnarray}
where $\mathcal{C}_{a} = 4 G^{2}/\kappa_{a}^{2}$. Crucially, the anti-damping does not scale with the gain (though we need $\kappa_{a} < \kappa$) leading to no limitation on the gain-bandwidth product for this system. 
\par
%
Fig.~\ref{Fig.:DPSAchar} depicts the relevant figures of merit for the directional PSA, calculated including the relevant next sideband contributions, i.e., counter rotating terms associated with $ \omega_0 = 2 \omega_b = 2\omega_a$ up to first order. Fig.~\ref{Fig.:DPSAchar}(a) shows that the auxiliary mode $Q_{a}$ needs to be sufficiently high in order to obtain useful directionality; this coincides with the results found for phase-preserving amplification with biharmonic Raman amplifiers (cf. Fig.~\ref{Fig.:ComparisonRaman}). Furthermore, as shown by Figs.~\ref{Fig.:DPSAchar}(c,d), for a given $Q_{a}$ having too large a $\kappa_{a}$ is also detrimental from the point of view of bandwidth and noise properties of such an amplifier. On the other hand, having a too small $\kappa_{a}/\kappa$ ratio is unfavorable for directionality, as evident from the calculation of the reverse gain. 
Our calculations show that the reverse gain is strongly suppressed for a large $\kappa_{a}/\kappa$. Note that this ratio is always limited to less than unity due to stability considerations; this constraint arises due to the feedback of the quadrature mixing term $\hat{P}_{2}\hat{U}$ in Eq.~(\ref{EqHPSAquad}). The desirable hierarchy of different frequency scales for stable directional operation, thus becomes $\kappa_{a} < \kappa  < \omega_{a}$. 
%
%
%
%
\section{Discussion}
\label{sec_discussion}
%
%
\subsection{Gain versus directionality bandwidth}
\label{sec_tradeoff}
%
The constraint of a constant gain-bandwidth product for biharmonic Raman amplifiers, discussed in Sec.~\ref{sec_Raman}, can be calculated using the expression for forward gain [cf. Eq.~(\ref{Eq.:Gain3modeNDPA})]. The forward gain is highest at resonance of the selected mode and decreases with increase in detuning as
\begin{eqnarray}
	\mathcal{G} [\omega] \underset{\mathcal{C}\rightarrow1^{-}}{=} \mathcal{G}_{0} \frac{f(\vartheta_{a})}{1 + \mathcal{G}_{0}\vartheta_{b}^{2}},
\label{Eq.:Gainvsdetuning}
\end{eqnarray}
where $\vartheta_{a} = \omega /\kappa_{a}, \vartheta_{b} = (\omega + \omega_{a})/\kappa_{b}$ for the USB amplifier and $\vartheta_{a,b} = \omega/\kappa_{a,b}$ for the RSB amplifier denote the respective reduced detunings. Here we have retained only leading order terms in $\vartheta_{b}$. Also $f(\vartheta_{a})$ denotes a polynomial function of $\vartheta_{a}$ which does not depend on the gain $\mathcal{G}_{0}$; hence it does not affect gain-bandwidth product and can be taken as unity for $\vartheta_{a} \ll 1$. Eq.~(\ref{Eq.:Gainvsdetuning}) allows us to write the instantaneous amplification bandwidth as
\begin{eqnarray}
	\Delta (\mathcal{G}_{0}) \approx \frac{2\kappa_{b}}{\mathcal{G}_{0}^{1/2}}.
\end{eqnarray}
This trade-off between maximum useful gain of an amplifier and the instantaneous bandwidth over which such a gain is realizable is universal in most parametric amplifier systems \cite{ArchanaJPC, Eichler2014}. The directional phase-sensitive amplifier, on the other hand, shows no gain-bandwidth tradeoff as shown in Fig.~\ref{Fig.:DPSAchar}(d) [cf. Eq.~(\ref{Eq.:FwdGainPSA})].
\par
The directionality of biharmonic amplifiers is also frequency-dependent; in order to quantify this, we introduce a directionality parameter [cf. Eq.~(\ref{Eq.:RevGainSidebandNDPA})]
\begin{eqnarray}
	d[\omega] \equiv 1 - \frac{\bar{\mathcal{G}}[\omega]}{\mathcal{G}[\omega]} = \frac{ 1 }{1 + \vartheta_{a}^{2}},
\label{Eq.:dparameter}
\end{eqnarray}
with $d =0$ corresponding to usual reciprocal or symmetric amplification and $d= 1$ corresponding to perfect nonreciprocity. This allows us to define the directionality bandwidth as
\begin{eqnarray}
	\Delta_{d} = 2\kappa_{a}.
\label{Eq.:dbandwidth}
\end{eqnarray}
Using arguments similar to those for the gain-bandwidth tradeoff, it is straightforward to see that $\Delta_{d}/2$ denotes the detuning from resonance at which the directionality parameter reduces to $d=0.5$ --- a value that corresponds to a 3 dB isolation between forward gain $\mathcal{G}[\omega]$ and reverse gain $\bar{\mathcal{G}}[\omega]$. Crucially, \emph{directionality bandwidth $\Delta_{d}$ is independent of amplification bandwidth $\Delta$}, a behavior generic to these three-mode directional amplifiers. This is further borne out by following observations: 
\begin{enumerate}[(i)]
  \setlength{\itemsep}{0pt}
  \setlength{\parskip}{0pt}
\item Eq.~(\ref{Eq.:dparameter}), and by consequence Eq.~(\ref{Eq.:dbandwidth}), hold true for both phase-preserving and phase-sensitive operations which show qualitatively different gain-bandwidth behavior
\item Unlike $\Delta$, $\Delta_{d}$ is not limited by the gain $\mathcal{G}_{0}$ or by the linewidth $\kappa_{b}$ of the amplified or deamplified mode, and strictly scales with the linewidth of the auxiliary mode alone. 
\end{enumerate}
\par
\begin{figure}[t!]
\centering
\includegraphics[width=1.0\columnwidth]{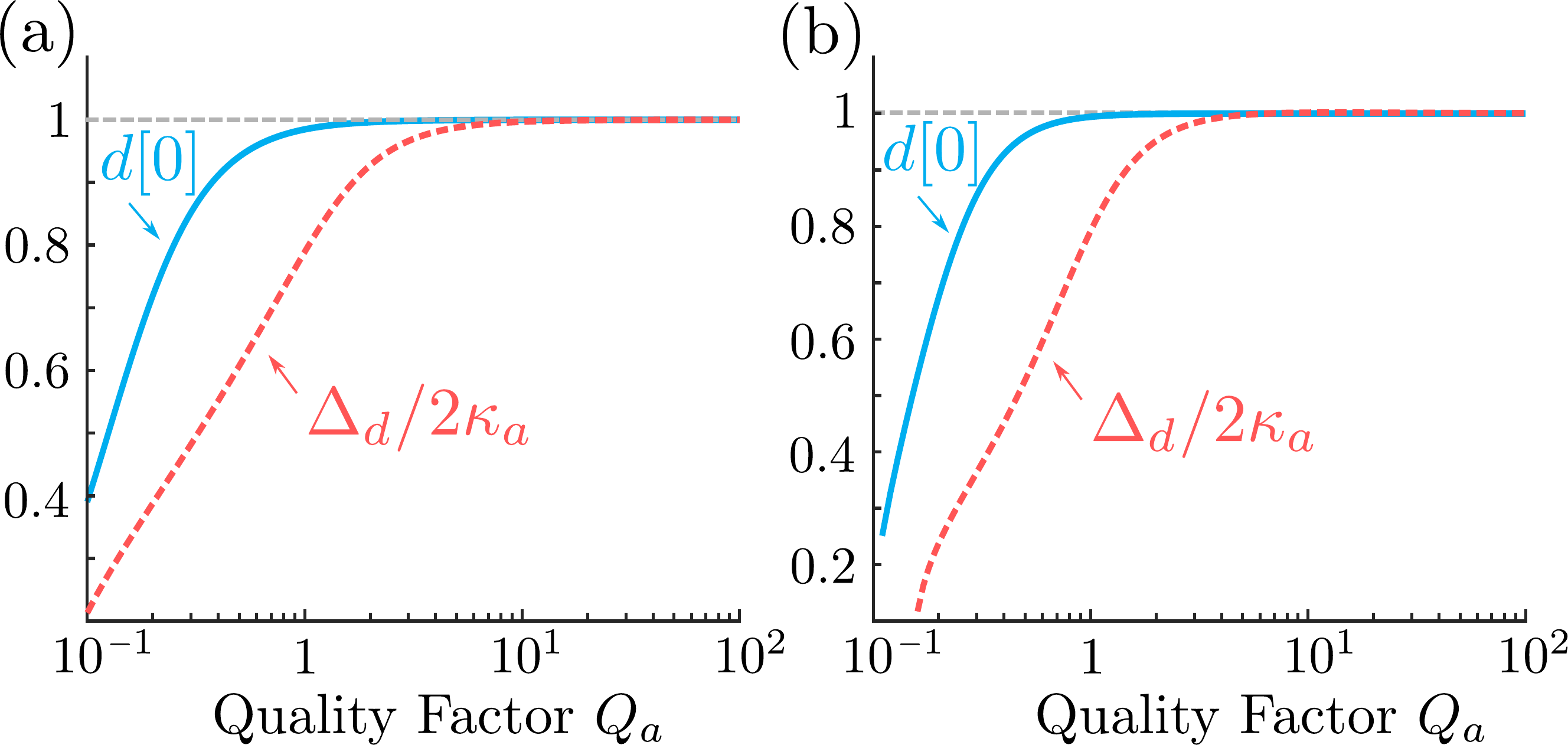}
	\caption{Variation of directionality parameter at resonance $d[0]$ (solid cyan) and directionality bandwidth $\Delta_{d}$ (dashed red), with the quality factor $Q_{a}$ of the auxiliary mode for (a) biharmonic Raman amplifiers, (b) directional phase-sensitive amplifier. Both plots were calculated for a forward gain of $\mathcal{G} = 20$ dB.}
\label{Fig.:Deltad}
\end{figure}

Therefore, in order to have directionality over a large bandwidth, it is essential for the auxiliary mode to have a proportionately large linewidth. However, as indicated by calculations including counter-rotating terms, too large $\kappa_{a}$ ($Q_{a} \rightarrow 0$) can cost net achievable directionality in these systems [Figs.~\ref{Fig.:ComparisonRaman} and \ref{Fig.:DPSAchar}(a)]. This is clear from Fig.~\ref{Fig.:Deltad} where we show an example calculation of the effect of counter-rotating terms in biharmonic Raman amplifiers, and the tradeoff they impose between the net achievable directionality $d$ and directionality bandwidth $\Delta_{d}$. 
In Fig.~\ref{Fig.:Deltad} the directionality parameter is evaluated on resonance ($\omega=0$). For the case of phase-preserving amplification we find the scaling
\begin{align}
d_{\rm CR}[0]  =&    \ \frac{   1 }{   1 +  \frac{1}{ 64 Q_a^2}  } ,
\end{align}
hence, for large $Q_a$ we have $d  \simeq 1$.
Note, that a directionality parameter of unity can only be achieved at resonance ($\omega=0$), the maximum attainable value of $d$ decreases quadratically with detuning, as per Eq.~(\ref{Eq.:dparameter}). 
In addition, effects of non-RWA corrections are slightly more pronounced at finite detunings. 
%
%
\par
Thus, while it is desirable to have the auxiliary mode in the steady state for stable device operation, it cannot be a very low-Q waveguide or a resistive load. A useful framework to distinguish the effects of auxiliary mode dynamics is to view this mode as an engineered reservoir, as elaborated in the following section.
%
%
\subsection{Connection to dissipation engineering}
\label{sec_dissengg}
%
%
It was recently shown that any factorisable coherent interaction can be rendered directional by balancing it with the corresponding dissipative interaction \cite{Metelmann2015}.
Dissipation is, therefore, the crucial element to obtain any directionality at all. This holds true as well for all biharmonic amplifier schemes proposed in this work. 
In this section we briefly sketch how this manifest itself, using the particular example of biharmonic Raman amplifiers. 
\par
We start out from the Hamiltonian in Eq.~(\ref{Eq.:Ham3mode}), describing a hopping (paramp) interaction between the upper (lower) sideband mode and the low frequency mode,
as well as mixing between the two sidebands  mediated by the second harmonic. The auxiliary mode $\hat{a}$ can be considered as the engineered reservoir that provides us with the desired dissipative interaction.   Elimination of this mode leads to the following coupled equations for the remaining modes 1 and 2 as
\begin{align}\label{Eq.:CouplingBRA}
 \hat d_{1}[\omega] \sim&            - i \left[ \widetilde G_2[\omega]   - i \frac{\Gamma[\omega]}{2}    \right] \hat d_2^{\dag}[\omega] ,
 \nonumber \\ 
 \hat d_{2}^{\dag}[\omega] \sim&     + i \left[ \widetilde G_2^{\ast}[\omega]   - i \frac{\Gamma[\omega]}{2}    \right] \hat d_{1}[\omega].
\end{align}
Here we consider the full frequency dependence of the auxiliary mode, and accordingly define the couplings
\begin{align}
 \widetilde G_2[\omega] &= G_2 e^{- i \alpha} +  \frac{   \frac{ \omega}{\kappa_a}  \Gamma_{0}  }{1 +  \frac{4\omega^2}{\kappa_{a}^2}  },
 \hspace{0.5cm}
  \Gamma [\omega]        =   \frac{\Gamma_{0}}{1 +  \frac{4\omega^2}{\kappa_{a}^2}  },
\end{align}
with $\Gamma_{0} = 4G_1^2/\kappa_a $. The first terms in Eqs.~(\ref{Eq.:CouplingBRA}) corresponds to a coherent interaction, they could be derived from an effective Hamiltonian of the form $\mathcal  H_{\rm coh} = \widetilde G_2[\omega] \hat d_1^{\dag} \hat d_2^{\dag} + h.c.$.  The same does not hold true for the second coupling terms,  they can never be derived from a coherent Hamiltonian. These terms correspond to a dissipative interaction mediated by the auxiliary mode, i.e., they could be derived from an effective  non-local dissipator
$\Gamma [\omega] \mathcal L  [\hat d_1 + \hat d_2^{\dag}  ] $ in a master equation \footnote{We use the definition $\mathcal L [\hat o] \hat \rho = \hat \hat o \hat \rho \hat o^{\dag}   - 1/2 \hat o^{\dag}\hat o \hat \rho  - 1/2 \hat \rho  \hat o^{\dag} \hat o  $.}. 
\par
The general condition of balancing a coherent interaction with its dissipative counterpart reported in Ref. \cite{Metelmann2015} translates into simply tuning the amplitude and the phase of the coherent coupling,
\begin{align}
 \big| \widetilde G_2[\omega] \big| = \frac{ \Gamma [\omega]}{2},
 \hspace{0.5cm}
  \arg( \widetilde G_2[\omega]) = \pm \frac{\pi}{2}.
\end{align}
Applying these conditions to Eq.~(\ref{Eq.:CouplingBRA}) renders the coupling between the two modes directional. This selective decoupling would not be possible without the dissipative interaction. In principle, the system could be rendered directional for every frequency with the above conditions. However, in an experiment the amplitude $G_{2}$ and the phase $\alpha$ will be fixed and the system is rendered completely directional at a single frequency, e.g., in the present frame this would be at resonance or $\omega = 0$. The frequency range around that frequency over which the reverse gain is suppressed is then determined by $\kappa_a$, i.e., the inverse memory time of the engineered reservoir as explained in the previous section. If we assume that this memory time is vanishingly small, i.e., that the auxiliary  mode $\hat{a}$ is strongly damped, we can treat the mode as a Markovian reservoir and the whole system can be modeled via a Lindblad master equation of the form
\begin{align}\label{Eq.:LdissNDPA}
\frac{d}{dt} \hat \rho =  -i \left[\mathcal H_{\rm coh} , \hat \rho\right] + 
                          \Gamma_0 \mathcal L \left[\hat d_1 + \hat d_2^{\dag} \right] \hat \rho.
\end{align}
Here the non-local dissipator describes a Markovian reservoir which absorbs excitation from mode $\hat d_1$ and emits it into $\hat d_2^{\dag}$, with a rate $\Gamma_0$. In the overdamped case the master equation sufficiently describes the system. However, for arbitrary damping $\kappa_a$  one has to only include the non-Markovian effects due to a finite life-time of the reservoir, i.e., the low-frequency mode.
%
%
\section{Conclusions}
\label{sec_conclusions}
%
We have studied different modalities of directional quantum-limited amplification realizable in a three-mode system pumped with a biharmonic pump. For optimal amplitude and phase difference between the two harmonics and appropriate choice of mode frequencies, such a system provides the minimal implementation of nonreciprocal photon transmission and amplification. 
We explicitly present the schemes  for a directional phase-preserving (both degenerate and non-degenerate) and phase-sensitive amplification.
The generality and minimality of our proposals should make it suitable for implementation in multiple platforms, such as superconducting qubits and opto- or electro-mechanical systems. Using pump harmonics can be particularly desirable in optical systems, where supplemented by second/sub harmonic generation, it can drastically reduce the resource overhead in nonreciprocal optical platforms. 
\par
We also evaluate full frequency-dependent forward and reverse gains, and the available bandwidth with each scheme. Our results show that there is a universal separation of parameters determining directionality and amplification bandwidths. In particular, the directionality bandwidth increases directly with the linewidth of the dissipative/auxiliary mode alone. On the other hand, the net magnitude of directionality is predicated on a modestly high quality factor for the engineered reservoir mode, required to suppress the deleterious counter-rotating contributions.
%
%
\section{Acknowledgements}
\label{sec_thanks}
%
The authors wish to thank Aashish Clerk, Michel Devoret, Leonardo Ranzani and John Teufel for useful discussions. 
\appendix
%
\section{Single-pump Raman amplifier}
\label{app_SRA}
%
\begin{figure}[h!]
\centering
 \includegraphics[width=0.9\columnwidth]{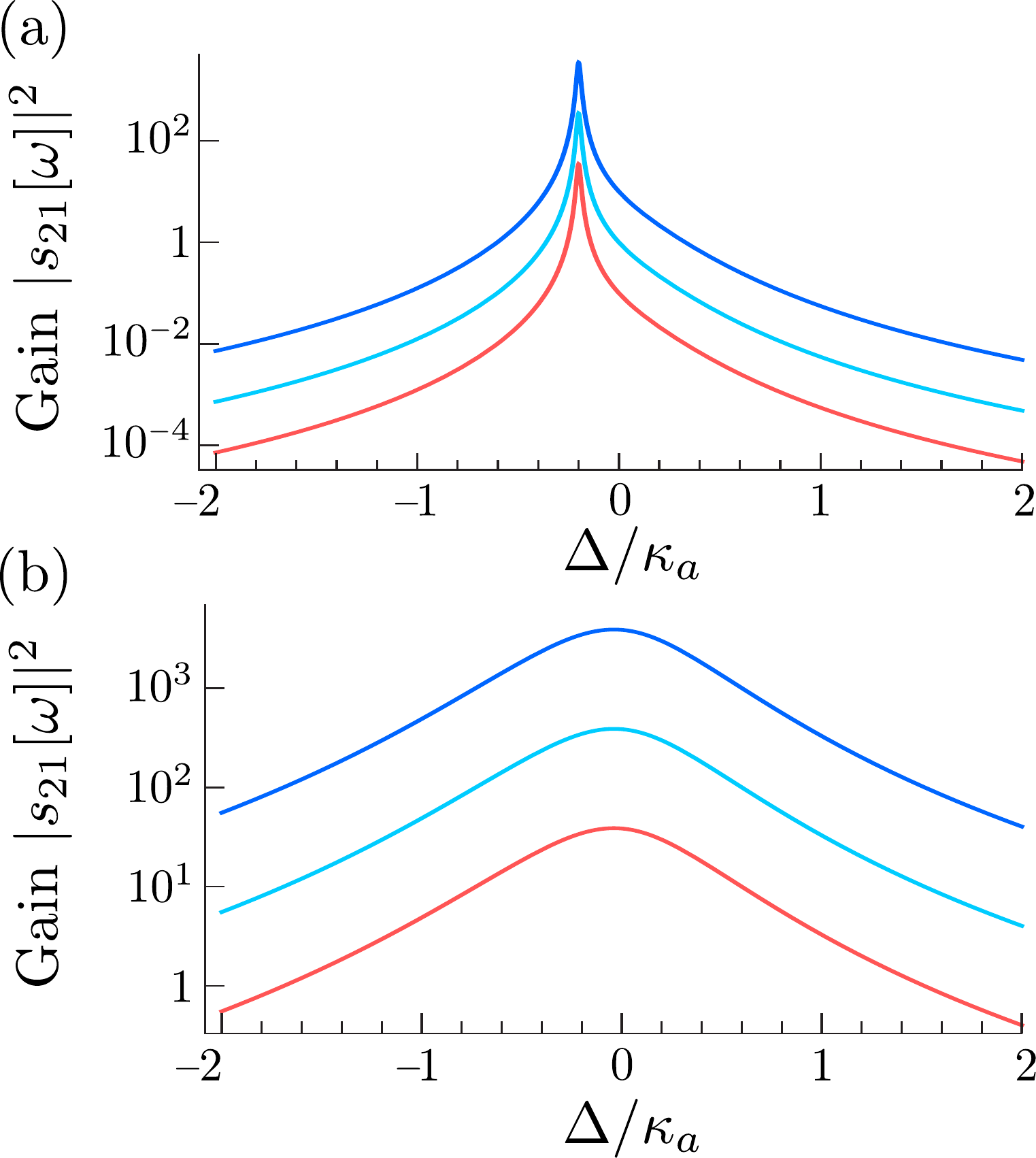}
	\caption{Gain profiles of single-pump Raman amplifier calculated for (a) $\omega_{a}/\kappa_{a} = 10, \; \omega_{a}/\kappa_{b} = 0.2$ and (b) $\omega_{a}/\kappa_{a} = 0.1, \; \omega_{a}/\kappa_{b} = 0.2$, with three values of cooperativity ${\mathcal{C} = 10 \; {\rm (red)},\; 100\; {\rm (cyan)},\;1000\; {\rm (blue)}}$. Note that here due to absence of any instability in the system, the cooperativity $\mathcal{C}$ does not need to be limited to a value less than unity.}%
\label{Fig.:SRA}
\end{figure}
We consider a single pump modulation $M(t) = G (e^{-i \omega_p t }+ e^{ i \omega_p t })$ for the Raman amplifier in the unresolved sideband regime [Fig.~\ref{Fig.:Raman}(a)], with $\omega_{p} = \omega_b$. Going into an interaction picture with respect to the free Hamiltonian, we obtain,
\begin{align}
 \hH' =  G_{1} \ (\hat a e^{-i \omega_{a} t } +\hat a^{\dag} e^{i \omega_{a} t }) 
	     (\hat b  + \hat b^{\dag}  ) + \hH_{\rm CR}+ \hH_{\rm bath}',
\label{Eq.:HamilSRA}
\end{align}
with the counter-rotating terms,
\begin{equation}
\hH_{\rm CR}= G_{1} (\hat b e^{-i 2 \omega_b t }+\hat b^{\dag} e^{ i 2 \omega_b t })(\hat a e^{-i \omega_{a} t } +\hat a^{\dag} e^{i \omega_{a} t }).
\end{equation}
Ignoring the counter-rotating terms under the assumption of a high-$Q_{a}$ for the low-frequency oscillator, i.e., $\kappa_a \ll \omega_a < \kappa_b$, we can approximate the zero-frequency scattering matrix as,
\begin{equation} 
 \mathbf{S}[0] \approx
\left(
\begin{array}{ccc}
	  \displaystyle     - 1
	& \displaystyle   2 i \sqrt{\mathcal C}  
	& \displaystyle   2 i \sqrt{\mathcal C} 
\\[0.2cm]
	  \displaystyle  2 i \sqrt{\mathcal C}
	& \displaystyle   \left( 2  \mathcal C -  1  \right)
	& \displaystyle  2   \mathcal C  
\\[0.2cm]
	  \displaystyle  -2 i \sqrt{\mathcal C}
	& \displaystyle   - 2  \mathcal C  
	& \displaystyle   \displaystyle    -\left( 2  \mathcal C +  1  \right) 
\\
\end{array}
\right),
\end{equation}
where ${\mathbf{\hat D}_{\rm out}[\omega] = \mathbf{S}[\omega]  \mathbf{\hat D}_{\rm in}[\omega]}$ with 
${\mathbf{\hat D}_{\rm in}[0] = [\hat a_{\rm in}^{\PD}[0], \hat b_{\rm in}^{\PD}[\omega_a],\hat b_{\rm in}^{\dag}[\omega_a]]^{\rm T}}$ and we have introduced the cooperativity $\mathcal{C}= 4 G_{1}^{2}/(\kappa_{a}\kappa_{b})$. Note that in the absence of the second harmonic, the scattering is reciprocal. The expression for frequency-dependent inter-modulation gain between modulation frequency $\omega_{a}$ and sidebands $\omega_{\pm}$, for this case, is given by
\begin{equation}
	|s_{21}[\omega]|^{2}= \frac{4 \mathcal{C}}
{\left[1 + \frac{4 \omega^{2}}{\kappa_{b}^{2}} \right] \left[1 + \frac{4 (\omega + \omega_{a})^{2}}{\kappa_{a}^{2}} \right]}.
\end{equation}
As shown in Fig. \ref{Fig.:SRA}, the bandwidth does not shrink with increasing gain unlike the usual parametric amplification schemes. Also, the bandwidth scales as ${\rm min}[\kappa_{a}, \kappa_{b}]$.
\par
The transmission power gain $|s_{21}[\omega]|^{2}$ is associated with a frequency conversion process, i.e., a signal injected at the low frequency mode will be upconverted to the lower sideband frequency.
However, the system works as an ampiflier without any frequency-conversion in reflection as well, this means that a signal injected on either sideband is reflected with amplitude gain $2\mathcal C - 1$. 
This realization of a phase-sensitive amplifier has been demonstrated experimentally \cite{John} and has a close connection to the dissipative amplifier discussed in Ref.~\cite{Metelmann2014}. 
This becomes obvious if we use the mapping defined in Eq.~(\ref{Eq.MappingUSB}) in Eq.~(\ref{Eq.:HamilSRA}) that yields an interaction,
\begin{align}
 \hH =&    G_1 \hat a  \left(   \hat b_{+}^{\dag} +   \hat b_{-}   \right)  + h.c.,
\end{align}
corresponding to a hopping (amplifier) interaction between the low frequency mode and the upper (lower) sideband.  
Elimination of the low frequency mode realizes a coupling between the modes which could not be achieved via a coherent interaction,
but rather by a nonlocal dissipator of the form $\mathcal L [\hat b_{+}^{\dag} + \hat b_{-} ] \hat \rho$.
\par
Clearly, the same 'dissipative' amplifier could be realized in the resolved sideband regime, i.e., with two independent modes $\hat d_1$ and $\hat d_2$. Based on such a three mode setup, a recent experiment showed that broadband amplification close to the quantum limit is possible \cite{ockeloen-korppi_low-noise_2016}. 

%
%
%
\section{Gainless circulation with biharmonic pump}
\label{app_circ}
%
%
The general three-mode interaction Hamiltonian in Eq.~(\ref{Eq.:3modeHamiltonian}) can be tuned to realize a gainless nonreciprocal transmission between different channels, namely a frequency circulator. This kind of system was recently by Ranzani and Co-workers \cite{Ranzani2014} and experimentally realized in a JPC setup \cite{Sliwa2015}. The later experiment required three pump modes which drive the hopping between the three modes of the JPC, in such a way that  one realizes the Hamiltonian
\begin{align} 
 \hH =&    G_1 \left( \hat a \hat d_1^{\dag}  + \hat a  \hat d_2^{\dag}    \right)  
	+  G_2 \hat d_1 \hat d_2^{\dag} e^{- i \alpha}   + h.c.
\end{align}
Such a three-mode interaction can be realized, using only one pump and its second harmonic by selecting the driving frequencies,
\begin{align}
 \omega_{P,1} =&  \omega_1 - \omega_a  \equiv 2 \omega_0,
\nonumber \\
\omega_{P,2} = &  \omega_2 - \omega_a  \equiv \omega_0,
\nonumber \\
\omega_{P,3} = &  \omega_1 - \omega_2   \equiv     \omega_0,
\end{align}
which implies $\omega_{a} + \omega_{1} = 2\omega_{2}$. For example, if one would work with the pump frequency $\omega_{0}/(2\pi) = 4$~GHz and the low frequency mode at $\omega_{a}/(2\pi)= 1$~GHz, one would need to have $\omega_{1}/(2\pi) = 9$~GHz and $\omega_2/(2\pi) = 5$~GHz for the remaining two oscillators. 
%
%
%

%
%
%
\end{document}